\documentclass{article}
\usepackage{spconf,amsmath,graphicx}

\usepackage{enumitem}
\setlist{nosep, leftmargin=14pt}

\usepackage{mwe} 

\usepackage{graphicx}
\usepackage{comment}
\usepackage{amssymb}

\usepackage{mathtools}
\usepackage{dsfont}

\newcommand{\ie}{i.e., }

\title{Integrating multiscale topology in digital pathology with pyramidal graph convolutional networks}

\name{
\begin{tabular}{@{}c@{}}
Victor Ibañez \qquad Przemyslaw Szostak \qquad Quincy Wong \qquad Konstanty Korski\\ Samaneh Abbasi-Sureshjani$^{\star}$ \qquad Alvaro Gomariz$^{\star}$\thanks{$^{\star}$ These two last authors have equal contributions}
\end{tabular}
}
\address{
F. Hoffmann-La Roche AG, Basel, Switzerland\\
}
\begin{document}
\ninept
\maketitle{}
\begin{abstract}
Graph convolutional networks (GCNs) have emerged as a powerful alternative to multiple instance learning with convolutional neural networks in digital pathology, offering superior handling of structural information across various spatial ranges --- a crucial aspect of learning from gigapixel H\&E-stained whole slide images (WSI).
However, graph message-passing algorithms often suffer from oversmoothing when aggregating a large neighborhood. Hence, effective modeling of multi-range interactions relies on the careful construction of the graph.
Our proposed multi-scale GCN (MS-GCN) tackles this issue by leveraging information across multiple magnification levels in WSIs.
MS-GCN enables the simultaneous modeling of long-range structural dependencies at lower magnifications and high-resolution cellular details at higher magnifications, akin to analysis pipelines usually conducted by pathologists.
The architecture's unique configuration allows for the concurrent modeling of structural patterns at lower magnifications and detailed cellular features at higher ones, while also quantifying the contribution of each magnification level to the prediction. 
Through testing on different datasets, MS-GCN demonstrates superior performance over existing single-magnification GCN methods. 
The enhancement in performance and interpretability afforded by our method holds promise for advancing computational pathology models, especially in tasks requiring extensive spatial context.
\end{abstract}

\section{Introduction}
\label{sec:intro}
Digital pathology is transforming with deep learning techniques, notably using H\&E stained images for various clinical tasks. The gigapixel scale of whole slide images (WSIs) presents a challenge for conventional Convolutional Neural Networks (CNNs) which typically require entire images to be loaded into memory for processing. Current strategies often involve dividing WSIs into smaller tiles and applying multiple instance learning (MIL), where each 'bag' of tiles inherits the slide's label for training purposes\cite{maron1997framework, gadermayr2022multiple}. While this approach has yielded success, it principally captures local contextual information and neglects the broader tissue and cellular interactions.

Graph-based representation of whole slide images (WSIs) facilitates the modeling of complex interactions within the tissue architecture. Recent advancements employ graph convolutional networks (GCNs) to analyze these graphs, leveraging message-passing algorithms to learn from vertices connected by edges~\cite{jaume2021histocartography,lu2022slidegraph,chen2021whole}. Notably, PatchGCN~\cite{chen2021whole} effectively predicts cancer survival outcomes by constructing graphs where each vertex corresponds to a WSI tile and edges link neighboring tiles. This approach allows for a hierarchical aggregation of information, encapsulating local morphological features within their larger structural context.

Graph Convolutional Networks (GCNs), such as PatchGCN, have enhanced local contextual analysis in whole slide images (WSIs).

Yet, the broader tissue context modeling remains a challenge. Message passing algorithms, essential for GCNs, often struggle to capture interactions between distant regions because of oversmoothing. Additionally, directly connecting far-away vertices to counter this can result in excessively large graphs, which impede the learning process.
We address this by harnessing the inherent multiscale nature of slide analysis—mimicking pathologists who shift from low to high magnifications for a holistic to detailed examination. 
Indeed, WSI images are acquired at different \emph{magnification} levels, but current digital pathology algorithms often neglect the lower ones, instead focusing on a $20X$ level to balance detail with computational feasibility~\cite{de2018automatic, levy2019pathflowai, chen2021whole}.
As a consequence, valuable information from other scales is often omitted. 
Some multiscale approaches exist, but they mostly use CNNs that handle magnifications in isolation before combination, for example by using two different networks for different magnifications \cite{van2021hooknet}, or with an attention-MIL approach on $5X$ and $10X$ tiles \cite{li2021multi}.
The only attempt to do so to date is based on a relational GCN that processes each magnification independently, and subsequently concatenates information from different magnification layers~\cite{bazargani2022multi}.
While this method reports good results, the contribution of single vertices to the prediction is no longer traceable, hence losing the benefit of interpretability of the magnification information. 

\begin{figure}[htb]
\includegraphics[width=\columnwidth]{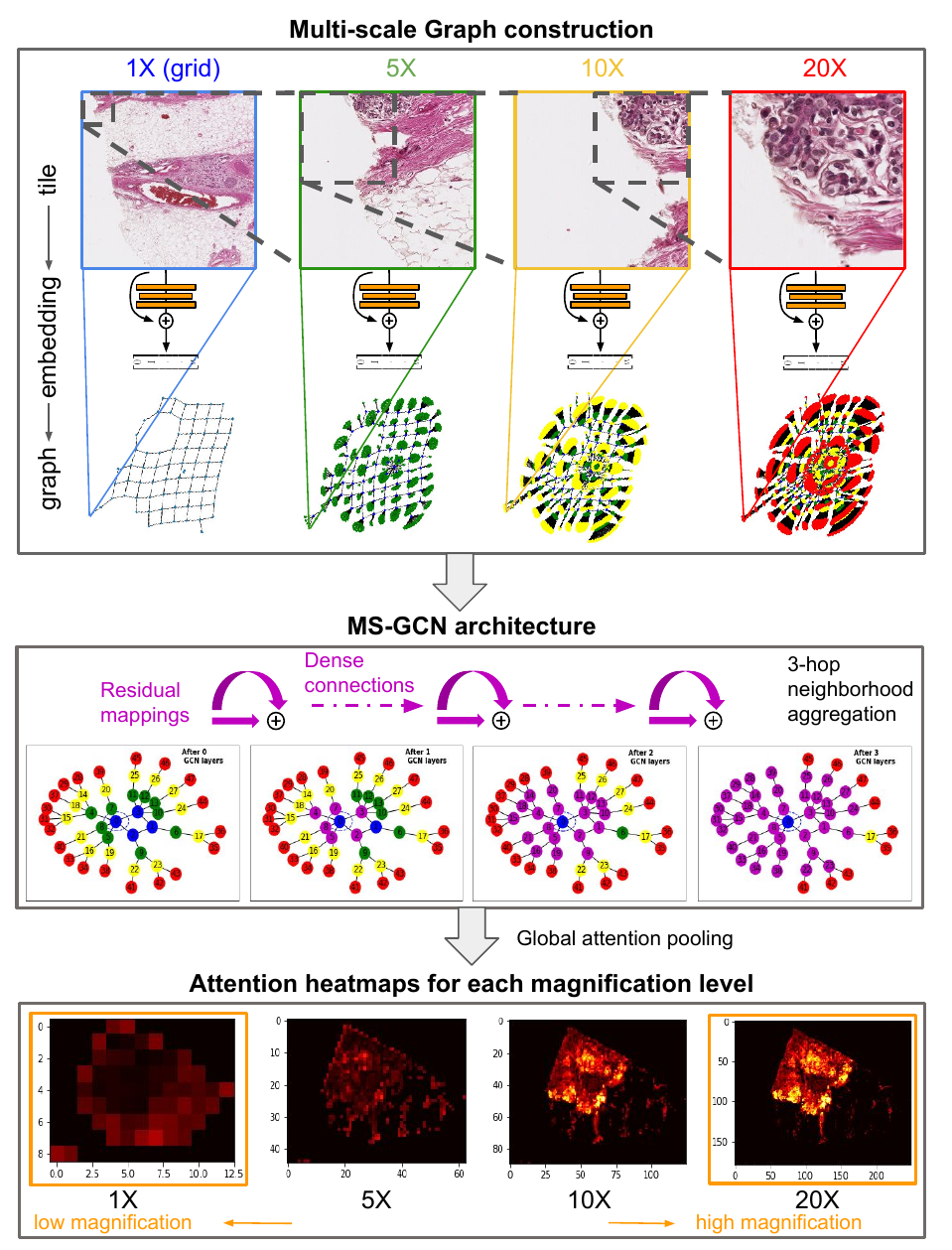}
\caption{Illustration of proposed MS-GCN method. (\textbf{top}) Higher magnification vertices (color coded) are connected to lower magnification ones recursively. Only the lowest resolution vertices are connected spatially among them. (\textbf{middle}) Message passage (magenta) algorithm in a graph following the color codes from top. (\textbf{bottom}) Magnification-specific attention heatmaps.
} \label{fig:overview}
\end{figure}

We herein propose a multi-scale graph convolutional neural network (MS-GCN) that builds upon PatchGCN by leveraging the information of different magnification levels in the graph construction. 
As illustrated in Fig.\ \ref{fig:overview}, MS-GCN can inherently separate the information into a spatial and a magnification part by ($i$) learning long-range structural information in the lowest magnification level where the tiles are connected, and ($ii$) adding pyramid graph connections from each low magnification tiles to their corresponding regions in higher magnification tiles recursively.
Hence, our approach integrates multiscale data into GCN to capture a comprehensive and interpretable pathology landscape, leading to a superior performance and interpretability of the different magnification features.

\section{Methods}

\subsection{Multi-scale graph construction}
We propose a multi-scale graph framework, where the graph $\mathcal{G}$ for a WSI is constructed across distinct magnification levels $m \in \{1 \ldots M\}$.
We denote $W^m$ the WSI corresponding to a resolution level $m$. Each $W^m$ is decomposed into different non-overlapping patches $\mathbf{X}^m_i \in W^m$.

The graph is defined as $\mathcal{G} = \{\mathcal{V}, \mathcal{E}\}$. $\mathcal{V}$ is the set of vertices $v_i,\, \forall i \in \{1 \ldots V\}$ centered on each of the patches $\mathbf{X}_i^m$. 
We denote a vertex $v_i^m$ when it is centered on a patch corresponding to a magnification $m$. $\mathcal{E}$ is the set of edges $e_{i,j},\, \forall i,j \in \{1 \ldots V\}$ representing the connectivity between them, \ie if $e_{i,j} \in \mathcal{E}$, then $e_{i}$ and $e_{j}$ are connected to each other.

For the proposed MS-GCN, we define the overall edge set $\mathcal{E}$ as a union of a spatial subset $\mathcal{E}^{\text{spatial}}$ and a magnification subset $\mathcal{E}^{\text{mag}}$.

\noindent{\textbf{Spatial:}} We define spatial relationships with $\mathcal{E}^{\text{spatial}}$ between adjacent vertices only in the lowest magnification level, \ie vertices $v_i^1$, such that long-range structural modelling can be effectively achieved.
4-connectivity is employed to link each vertex to its nearest neighbors if they exist, only for vertices $v^1_{i}$:
\begin{equation*}
\mathcal{E}^{\text{spatial}} = \{e_{i,j} \,|\, v^{1}_i \text{ and } v^{1}_j \text{ are Von Neumann neighbors}\}   
\end{equation*}

\noindent{\textbf{Magnification:}} Finer details are learned from higher magnification vertices that are recursively connected to low magnification vertices $v^1_i$ through $\mathcal{E}^{\text{mag}}$.
We construct hierarchical connections across different magnification levels to form a multi-scale graph. Each vertex $v^{m}_i$ is linked to $n$ vertices $v^{m+1}_{j_1} \ldots, v^{m+1}_{j_n}$ at a higher magnification level $m+1$ if they exist, where these vertices corresond to patches that map to the same tissue region as the patch from which $v^m_i$ was derived. In this way, $n \leq \left( \frac{\text{mag}(m+1)}{\text{mag}(m)} \right)^2$, where the function $\text{mag}(m)$ maps the magnification index $m$ to its corresponding actual magnification level.
This architecture ensures that inter-magnification connections are present exclusively between vertices corresponding to the same underlying tissue features but at different resolutions, as defined by:
\begin{equation*}
\mathcal{E}^{\text{mag}} = \{e_{i,j} \,|\, v^{m}_i \text{ and } v^{m+1}_j \text{ map to the same region}\}
\end{equation*}

For the actual computing with the GCN operator, we map the vertex connectivity into an adjacency matrix $\mathbf{A}$ and store the vertex features in the matrix $\mathbf{H}$.
The adjacency matrix $\mathbf{A}$ is defined as $\mathbf{A} \in \{0,1\}^{V \times V}$, where $\mathbf{A}_{i,j} = 1$ if there exists an edge $e_{i,j}$ ; otherwise, $\mathbf{A}_{i,j} = 0$.
Each vertex is associated to a feature vector $\mathbf{h}^m_i$ that corresponds to a patch $\mathbf{X}_i^m$. A feature vector $\mathbf{h}^m_i \in \mathbb{R}^{1024}$ is calculated as $\mathbf{h}^m_i = F_\mathrm{fv}(\mathbf{X}^m_i)$, where $F_\mathrm{fv}$ is a ResNet-50 backbone pretrained on ImageNet and truncated after the third layer, which acts as feature extractor. From all the feature vectors $\mathbf{h}^m_i$ we build the vertex feature matrix $\mathbf{H} \in \mathbb{R}^{V\times1024}$.
These vertex feature matrix and adjacency matrix are used to build the graph as $\mathcal{G} = (\mathbf{H},\mathbf{A})$.

\subsection{MS-GCN architecture}
For MS-GCN we adopt the PatchGCN~\cite{chen2021whole} architecture to work on our multiscale approach. 
In this framework, the specialized graph representation learning operator $\mathcal{F}_{\mathrm{GEN}}$~\cite{li2019deepgcns,kipf2016semi} is used to learn from graphs, together with a residual mapping~\cite{li2020deepergcn}.
Multiple $\mathcal{F}_{\mathrm{GEN}}$ layers are stacked to learn global-level morphological features in the multi-scale graph as:
\begin{equation*}
\mathcal{G}^{l+1}=\mathcal{F}_{\mathrm{GEN}}(\mathcal{G}^{l};\phi,\rho,\zeta) + \mathcal{G}^{l}
\end{equation*}
where $\phi$, $\rho$ and $\zeta$ are all differentiable and learnable functions, representing the typical components of GCNs with message construction, message aggregation and vertex update, respectively.

We use dense connections from the output of every layer to the last hidden layer. We define the number of layers as $L=M-1$ to aggregate a $L$-hop neighborhood. 
This structure, along with the design of $\mathcal{G}$, ensures that embeddings at the highest magnification level ($\mathbf{h}_j^M$) aggregate information across all magnifications specific to their region.
Conversely, embeddings at the lowest magnification level ($\mathbf{h}^1_i$) aggregate not just multi-level information but also incorporate embeddings from spatially adjacent vertices in the grid. This approach enables the effective capture of long-range spatial patterns, essential for a comprehensive analysis in digital pathology.

Finally, a global attention-based pooling layer \cite{ilse2018attention} is included that computes a weighted sum of all vertex features in the graph. In our setting, this leads to a multi-scale embedding $\mathbf{z} \in \mathbb{R}^{1 \times V}$.

\subsection{Influence scores for multiscale interpretability}
\label{sec:influence}
We interpret the influence of each magnification on the final prediction by taking the resulting attention embedding or \emph{attention heatmap} $\mathbf{z}$ and splitting it into $M$ different embeddings corresponding to each of the magnifications $m$. 
Since each of the entries in $\mathbf{z}$ corresponds to a vertex $v \in V$ that is associated with a magnification $m$, we can then define $\mathbf{z}^m$ as the subset of $\mathbf{z}$ with entries corresponding to $m$.

To evaluate the contribution of each magnification level across the dataset, we introduce the \emph{average scaled influence score}. This metric is computed by first determining the median of $\mathbf{z}^m$ separately for each magnification level $m$ across all WSIs. These median values are then normalized to fall between 0 and 1. This normalization allows for a consistent and comparative assessment of the relative impact of different magnification levels on the overall analysis.

\section{Results and Discussion}
\subsection{Dataset and implementation details}
Our evaluation leverages four distinct datasets, chosen for their heterogeneity, to robustly validate our MS-GCN model:

\noindent{\bf Breast cancer I:} 1'937 WSIs from an internal breast cancer dataset with imbalanced classes tumor grade I and tumor grade III (548 and 1'289 WSIs respectively). 
This data is stratified based on multiple clinical variables.
Tumor grade II is neglected because of the gradual nature of the grading system.

\noindent{\bf Breast cancer II:} 1'126 WSIs with 30 grade I and 1'096 grade III samples, providing a parallel dataset to breast cancer I.

\noindent{\bf PANDA:} Imbalanced prostate cancer data from the PANDA challenge \cite{bulten2022artificial}, formed by 10'658 biopsies with six classes: no cancer (2'880 WSIs) and grades in ISUP scale from 1 to 5 (2'663, 1'338, 1'232, 1'243, and 1'212 WSIs respectively). 
We waive exhaustive pre-processing steps commonly employed in the challenge benchmarking, since we use this data to compare two methods independently of the challenge.

\noindent{\bf Tissue structure:} In-house dataset of 1'123 WSIs. 
It comprises 561 biopsies, typically small tissue extractions, and 562 resections, which are larger surgical tissue removals.
This dataset is utilized to validate our hypothesis about magnification-specific information discernment.


Since PANDA includes multiple classes, the categorical crossentropy loss is used for training and the Quatratic Weighted Kappa (QWK) score for evaluation, while the other datasets use the binary crossentropy loss and AUROC.
All datasets are randomly split into 80\% training and 20\% testing.

To build the graph $\mathcal{G}$, patches $\mathbf{X}_i^m$ of size $256 \times 256$ are extracted from WSIs utilizing four magnification levels: $1X$,$5X$,$10X$,$20X$. 
A quality control segmentation algorithm is employed to only consider tiles containing tissue. 
The MS-GCN architecture is implemented using PyTorch Geometric and includes three $\mathcal{F}_{\mathrm{GEN}}$ layers.
Hyperparameter tuning is refined through 5-fold cross-validation, selecting the best performer across learning rates from $2 \times 10^{-3}$ to $5 \times 10^{-6}$, hidden dimensions from 32 to 128, dropout from 0 to 0.5, and batch sizes from 1 to 32.
Weight  decay regularizer is set to 5\% of the learning rate.
The models are trained on the full data from all folds for test set predictions, with an additional quantitative assessment of magnification-specific attention score contributions.


\subsection{Evaluating the performance of MS-GCN}
The evaluation of MS-GCN is carried out using two distinct datasets for breast cancer tumor grading, and the PANDA dataset, a benchmark for prostate cancer grading. These datasets present classification challenges that differ fundamentally in their pathological evaluation criteria. For instance, prostate cancer grading hinges on discerning architectural patterns within the tissue, potentially requiring less magnification diversity. Breast cancer grading, however, demands a nuanced analysis that considers both low-magnification architectural features and high-magnification cellular details.

The results, as illustrated in Table \ref{tab:results}, indicate that while MS-GCN surpasses Patch-GCN across all datasets, it achieves a notably higher margin of improvement on the breast cancer datasets. This performance enhancement aligns with our hypothesis that breast cancer grading benefits substantially from a multiscale approach, which can integrate and interpret information from both macrostructures and microdetails within the tissue. 
Furthermore, PANDA only contains biopsy WSIs, which are smaller in size than the resection WSIs available in the breast cancer datasets. This factor may also be limiting the performance of our approach in PANDA given the limited spatial reach. 


\begin{table}[tb]
\caption{Classification performance for Patch-GCN and MS-GCN.}
\resizebox{\linewidth}{!}{%
\centering
\begin{tabular}{cccc} 
\hline
Model & Dataset & Score & Performance \\
\hline\hline
Patch-GCN & breast cancer I & AUROC & 0.86\\
\textbf{MS-GCN (ours)} & breast cancer I & AUROC & \textbf{0.89}\\
\hline
Patch-GCN & breast cancer II & AUROC & 0.73 \\
\textbf{MS-GCN (ours)} &  breast cancer II & AUROC & \textbf{0.78}\\
\hline
Patch-GCN & PANDA & QWK &  0.84\\
\textbf{MS-GCN (ours)} &  PANDA & QWK & \textbf{0.85}\\
\hline
\end{tabular}
}
\label{tab:results}
\end{table}

\subsection{Discriminating tissue level information}
The discriminative power of MS-GCN across scales is evaluated on a straightforward task: distinguishing between tissue resections and biopsies within a breast cancer dataset. This classification leverages predominantly the tissue architecture evident at lower magnifications. With a AUROC score of 0.99 achieved by MS-GCN, our results substantiate the premise that information from varying magnification levels can be effectively segregated, in this case underscoring the proficiency of MS-GCN in identifying distinct tissue structures.

\subsection{Interpretability of multi-scale graph features}
Beyond its robust performance, a salient feature of MS-GCN is the interpretability of scale-specific feature contributions, facilitated by the \emph{average scaled influence scores} detailed in Section \ref{sec:influence}.

\begin{figure}[htb]
    \centering
    \includegraphics[width=1.\linewidth]{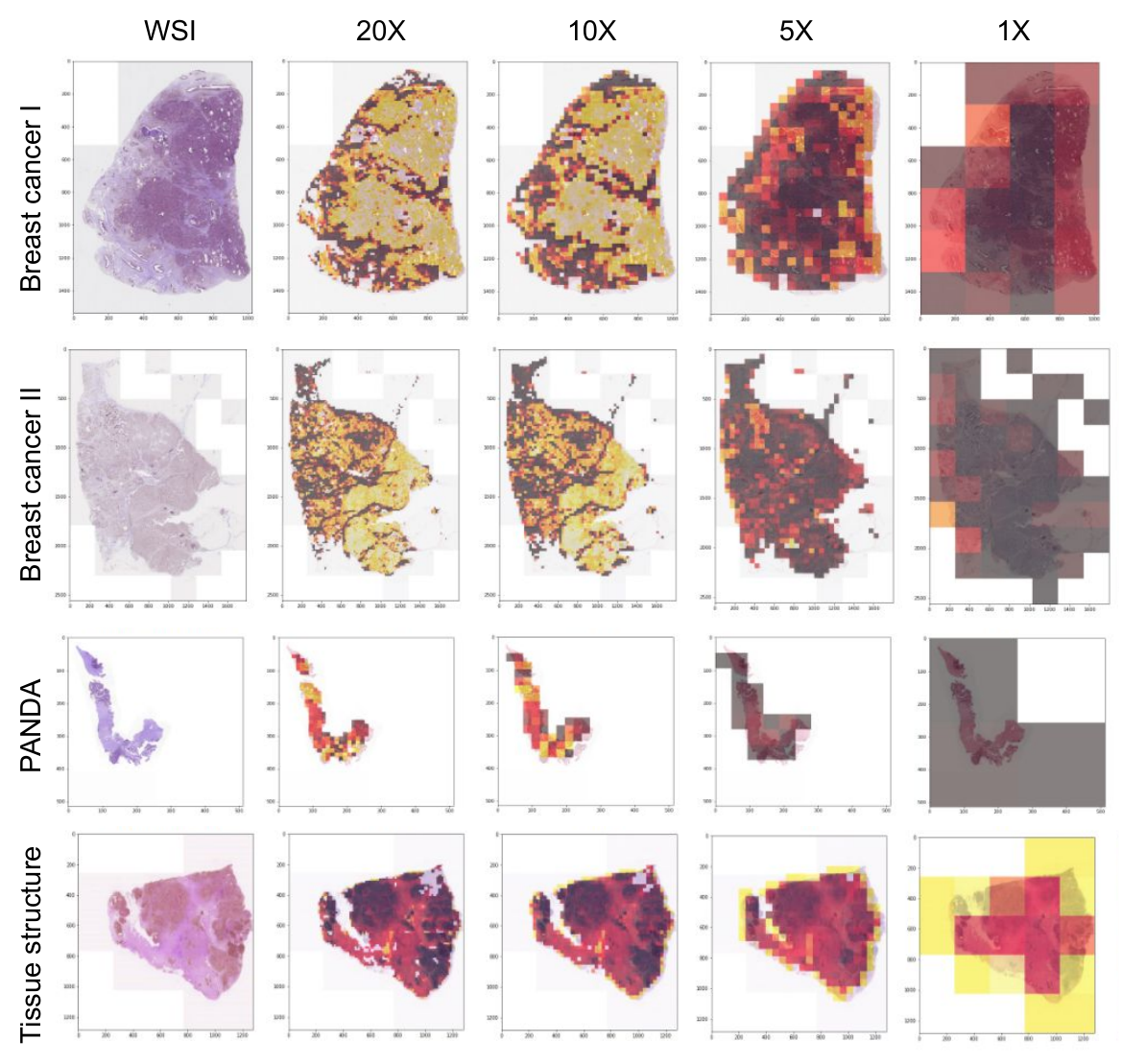}
    \caption{Attention heatmaps for the different datasets (rows). The columns show the raw WSI as well as the attention heatmaps at different magnifications as overlays on the WSI. 
    }
    \label{fig:attention}
\end{figure}

According to Table \ref{tab:influence}, for the breast cancer and PANDA datasets, the higher magnifications (10X and 20X) exert greater influence compared to the lower ones (1X and 5X). This pattern underscores the critical role of cellular-level features in cancer prediction over structural tissue characteristics visible at lower magnifications. Figure \ref{fig:attention} corroborates this, revealing the model's focus on cell-dense areas at $10X$ and $20X$ magnifications.

\begin{table}[htb]
\caption{Average scaled influence scores.}
\resizebox{\linewidth}{!}{%
\centering
\begin{tabular}{ccccc} 
\hline
Dataset & $m=1$ & $m=5$ & $m=10$ & $m=20$ \\
\hline\hline
breast cancer I & 0.04 & 0.09 & 0.42 & 0.45 \\
\hline
breast cancer II & 0.09 & 0.18 & 0.30 & 0.43\\
\hline
PANDA & 0.10 & 0.08 & 0.31 & 0.51 \\
\hline
Tissue structure & 0.42 & 0.26 & 0.17 & 0.15 \\
\hline
\end{tabular}
}
\label{tab:influence}
\end{table}

Conversely, the tissue structure dataset, which differentiates between biopsies and resections, exhibits a reversed trend, with $1X$ and $5X$ magnifications garnering significant attention. Visual analysis of the $1X$ attention heatmap in Fig.\ \ref{fig:attention} indicates a preference for border tiles.


\section{Conclusion}
Embracing multiple magnifications through our MS-GCN reflects the pathologist's approach to navigating the vast landscape of a WSI. Our research underscores the value of integrating lower magnifications, achieving superior performance over models employing graphs restricted to high-resolution analysis alone. The construction of a multiscale graph not only leverages the distinct features at each magnification level but also augments interpretability, mirroring the nuanced examination by experts. 

Integrating multiple magnification levels introduces a notable limitation:  increased computational demands, potentially impacting the method's scalability and broad applicability. Moving forward, benchmarking MS-GCN against state-of-the-art models across varied datasets, including different types of cancer, will be crucial to validate its effectiveness. Moreover, exploring diverse magnification combinations with GCNs that can model spatial dependencies at different ranges, together with  detailed analysis of influence scores across various datasets may unlock deeper understanding of morphological signatures.

\section{Compliance with Ethical Standards}
This research study was conducted using historical data from a clinical trial where patients gave consent for their data to be used in future research. All data is anonymised so individual patients cannot be identified, and no clinically relevant information was produced during this research.

\section{Acknowledgments}
F. Hoffmann-La Roche Ltd. provided support for the study and participated in the study design; conducting the study; and data collection, management, and interpretation. All authors were employees at this company at the time of the study. 

\bibliographystyle{IEEEbib}
\bibliography{refs}

\end{document}